\documentclass[a4paper,12pt]{article}

\usepackage{times}
\usepackage{graphicx}
\usepackage{subcaption}
\usepackage{multirow}
\usepackage{balance}
\usepackage{siunitx}
\usepackage{moreverb,url}
\usepackage{natbib}
\usepackage[colorlinks,bookmarksopen,bookmarksnumbered,citecolor=red,urlcolor=red]{hyperref}
\usepackage{geometry}
\usepackage{cite}
\title{Globus Service Enhancements for Exascale Applications and Facilities}

\author{
  Weijian Zheng\thanks{Argonne National Laboratory, Lemont, Illinois, USA}
  \and Jack Kordas\thanks{University of Chicago, Chicago, Illinois, USA}
  \and Tyler J. Skluzacek\thanks{Oak Ridge National Laboratory, Oak Ridge, Tennessee, USA}
  \and Raj Kettimuthu\footnotemark[1]
  \and Ian Foster\footnotemark[1]
}

\date{}

\begin{document}

\maketitle

\begin{abstract}
Many extreme-scale applications require the movement of large quantities of data to, from, and among leadership computing facilities, as well as other scientific facilities and the home institutions of facility users. These applications, particularly when leadership computing facilities are involved, can touch upon edge cases (e.g., terabyte files) that had not been a focus of previous Globus optimization work, which had emphasized rather the movement of many smaller (megabyte to gigabyte) files. We report here on how automated client-driven chunking can be used to accelerate both the movement of large files and the integrity checking operations that have proven to be essential for large data transfers. We present detailed performance studies that provide insights into the benefits of these modifications in a range of file transfer scenarios.
\end{abstract}

\section{Introduction}

Modern science applications often involve the processing of massive data from a wide spectrum of sources, encompassing both experimental and observational facilities, such as synchrotron light sources \citep{liu2021bridging} and telescopes, and supercomputers, as in climate science \citep{reichstein2019deep} and cosmology \citep{heitmann2019hacc}.
Workflows underpinning these science applications must be able to deliver these data to processing resources that best suit each applications' scale, timeliness, and hardware requirements. 
When data producers are remote from consumers, as they often are, these workflows must be able to transfer large data across high-bandwidth networks.
Reliable and rapid wide area data movement thus becomes a vital element of exascale computing systems \citep{alexander2020exascale}. 

The sharing of even extremely large data among geographically distributed resources and researchers has become increasingly feasible thanks to the widespread availability of high-bandwidth networks, the deployment of specialized network architectures \citep{dart2013science}, and the development of specialized data transfer protocols and services, notably the GridFTP protocol \citep{allcock2005globus} and the Globus service \citep{chard2016globus} which is widely used to manage GridFTP-based file transfers. 
Thus it has become commonplace to transfer petabytes over networks such as Internet2 and the U.S. Department of Energy's ESnet \citep{kettimuthu2018transferring,esgf_replication}.
However, emerging applications pose new challenges relating to the transfer of small numbers of extremely large (e.g., multi-TB) files, a workload for which Globus has not been optimized.
Here, we report on work that tackles these challenges by developing and evaluating enhancements to Globus for such transfers.
These enhancements focus on enabling the partitioning of large files, during a transfer, into many \textit{chunks} that can be transmitted concurrently. We show that this chunking can accelerate both data transfer and integrity checking operations performed as part of a transfer.

\section{Background}

\subsection{Scientific Data Transfer Infrastructure}

Modern scientific computing environments employ specialized data transfer infrastructure designed to maximize the speed achievable when moving data among geographically distributed storage systems.
These infrastructures typically combine a high-speed network; high-speed parallel file systems; a network and data transfer node (DTN) architecture to remove barriers to the rapid transfer of data over the network to/from file systems; and the Globus service and agents to achieve high-speed data movement. 

\begin{figure*}
    \centering
     \includegraphics[width=\textwidth]{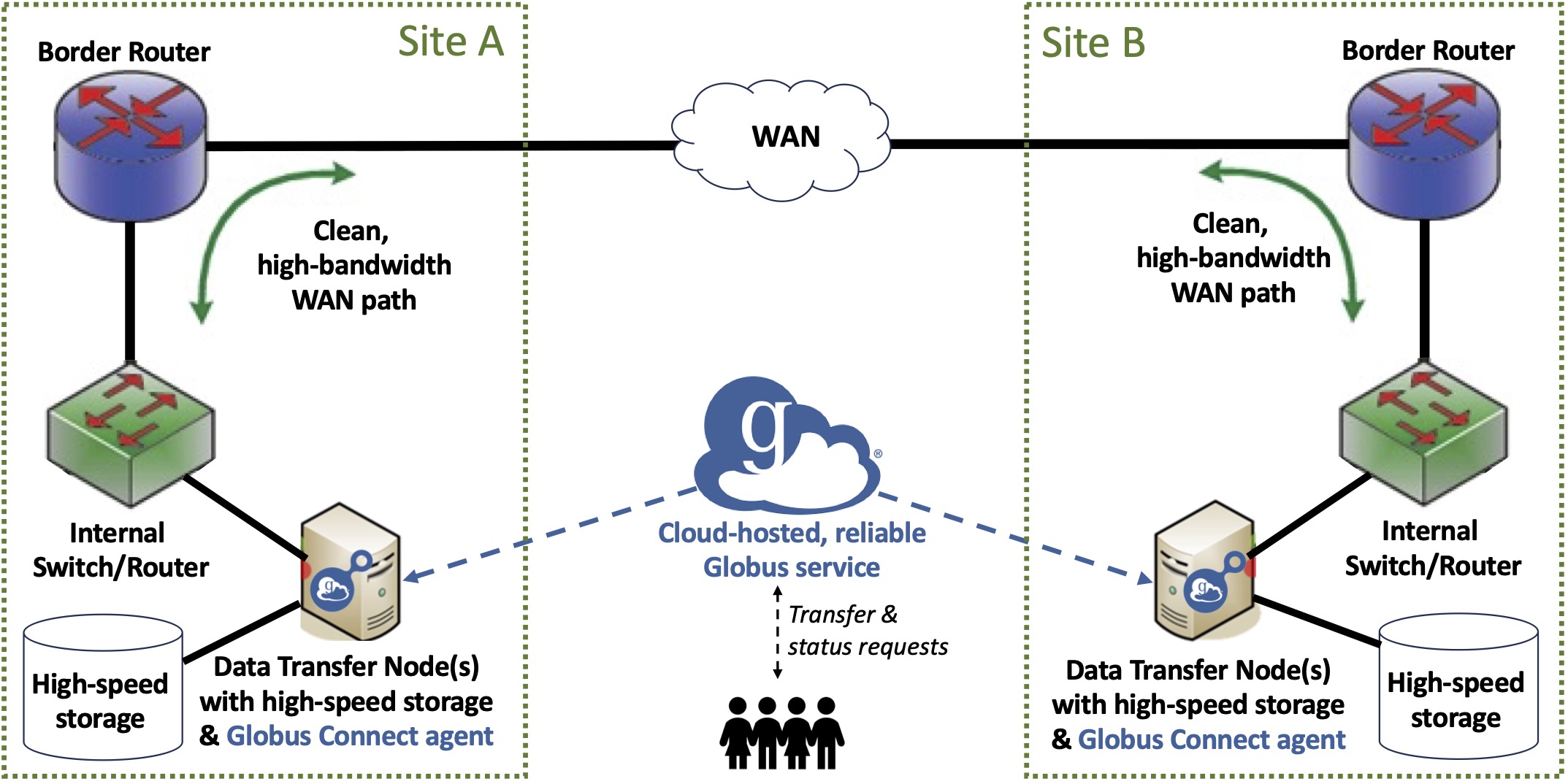}
    \caption{A modern data transfer infrastructure connects high-speed storage to wide area networks via a clean, high-bandwidth network path with one or more data transfer nodes (DTNs) hosting \textbf{Globus Connect agents}.
    The cloud-hosted \textbf{Globus service} acts as a client to the Globus Connect agents, instructing them to perform file transfers in response to user requests.
    }
    \label{fig:dtns}
\end{figure*}

Contemporary science networks such as ESnet connect research institutions at 100 Gb/s or higher rates.
These networks are optimized for high-speed, reliable packet transport.
At the end of these networks are typically high-performance parallel file systems such as Lustre (of which more below) that incorporate large degrees of internal parallelism to achieve high I/O rates. 
The elements that sit in between the external network and the file system play a crucial role in enabling high-speed data transfers.
Two essential elements \citep{dart2013science} are a clean network path to the external networks (without, for example, firewalls) and one or more 
Data Transfer Nodes (DTNs) configured to drive transfers at high speeds.
DTNs are specialized
servers configured specifically for efficient, high-speed data transfers. Specifically, they are equipped
with high-performance network interfaces, often 10 Gigabit Ethernet or higher; are configured
to optimize the data path to minimize latency and maximize throughput; are typically connected
directly to ESnet to take full advantage of its high-speed capabilities; are also connected directly
to high-speed storage; and run Globus Connect servers to handle large datasets, support reliable
multi-stream file transfers, and manage security, access control, and logging.

DTNs typically sit outside any corporate firewall(s) so that data can move between wide area network and HPC storage without interference.
To ensure that this configuration does not create security exposures, DTNs 
are configured with security measures tailored to the protection of data during transfer. This is particularly important when handling sensitive or proprietary information.

\subsection{GridFTP and Globus}

GridFTP \citep{allcock2003gridftp} is an extension of the standard File Transfer Protocol (FTP) designed specifically for high-performance, secure data transfer tasks. 
Its most widely employed implementation is that included in Globus \citep{allcock2001protocols,chard2016globus}, and in this brief summary we focus our discussion on that instantiation \citep{allcock2005globus}.

\begin{figure*}
    \centering
    \includegraphics[width=\textwidth]{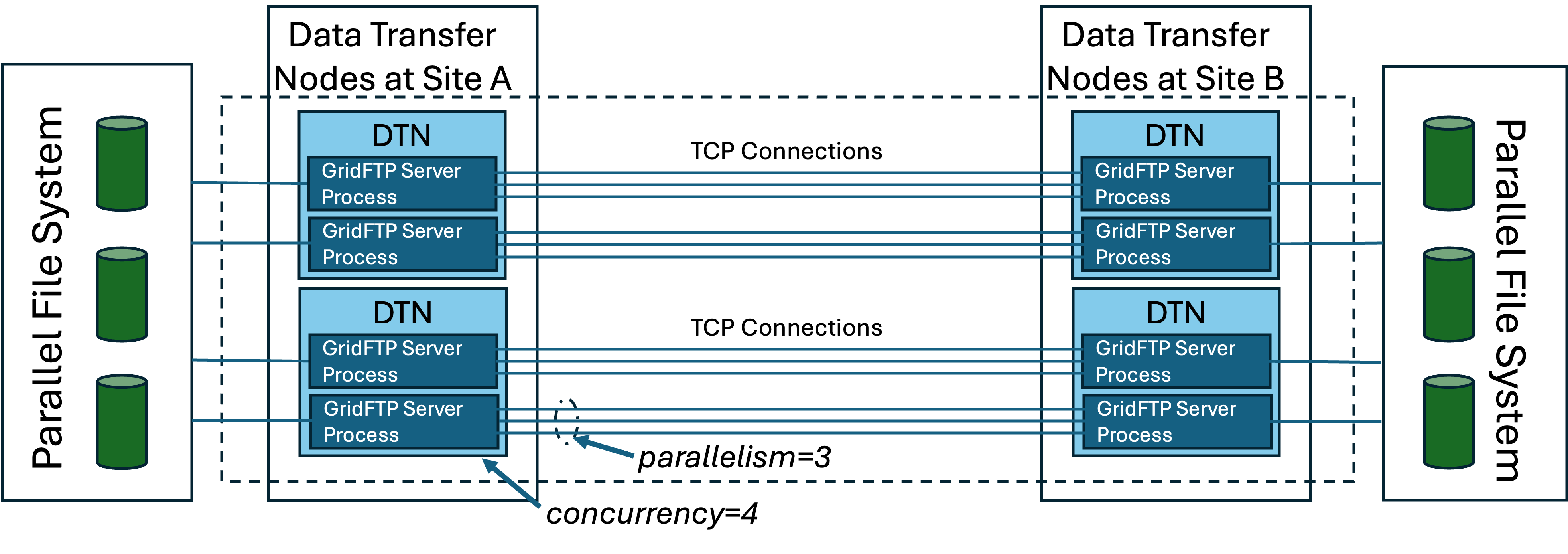}
    \caption{Concurrency (multiple data movers) and parallelism (multiple TCP connections) as implemented in Globus GridFTP.
    }
    \label{fig:globus-concurrency}
\end{figure*}

\begin{figure}[hb]
    \centering
    \includegraphics[width=\textwidth]{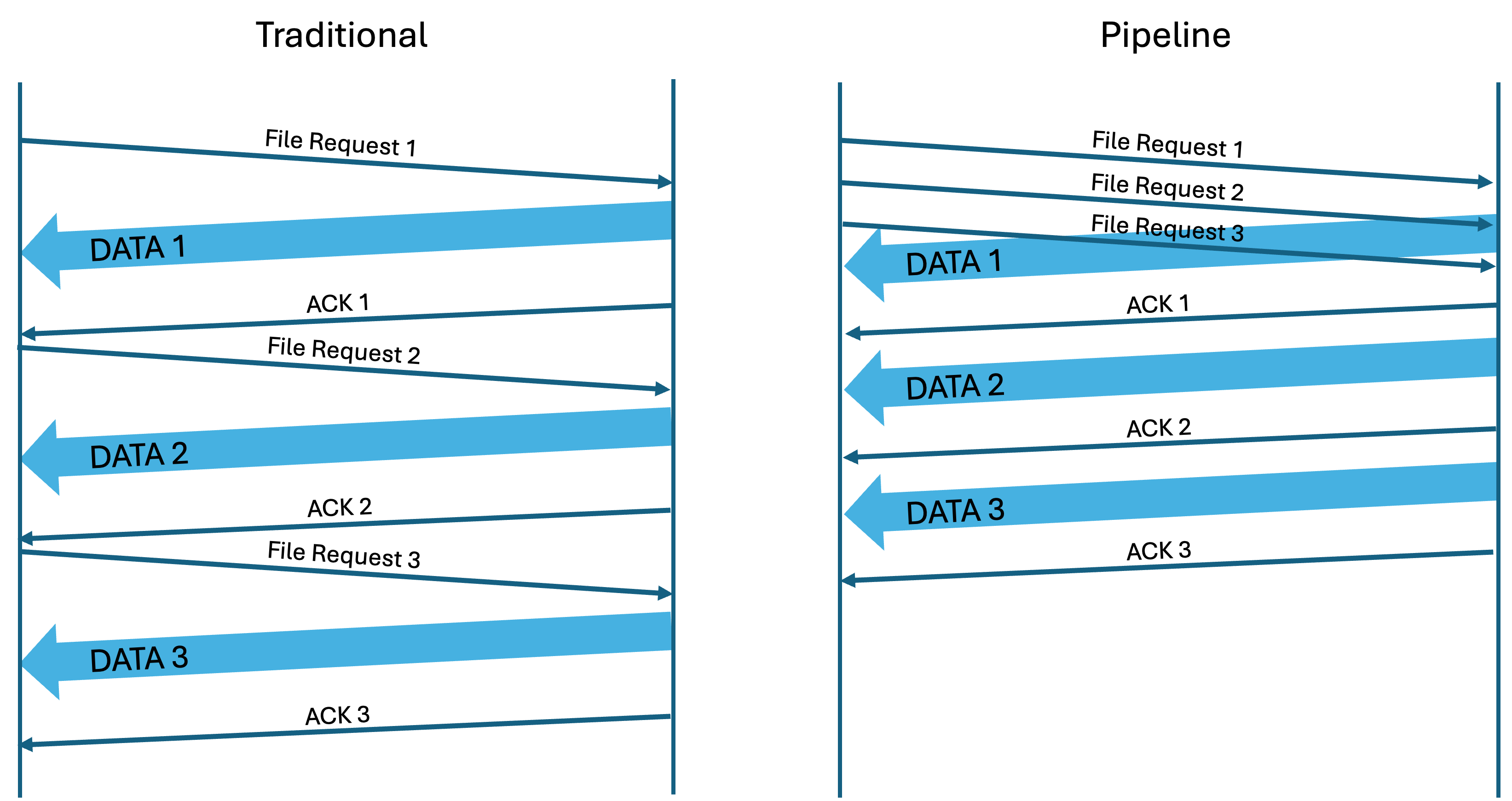}
    \caption{Pipelining in Globus GridFTP.
    Delays due to waiting for acknowledgements (left) are reduced by sending multiple requests at once (right).
    }
    \label{fig:globus-pipelining}
\end{figure}

The GridFTP protocol and its Globus implementation incorporate a variety of features designed for high-speed, reliable, and secure file transfer. 
For performance, Globus GridFTP employs \textit{parallelism}, whereby a single data mover employs multiple concurrent connections to transfer different parts of a single file, and \textit{concurrency}, whereby multiple data movers communicate different files: see \autoref{fig:globus-concurrency}.
It also employs \textit{pipelining}, in which multiple file transfer requests can be in flight without acknowledgments (see \autoref{fig:globus-pipelining}), and can handle third-party transfers, whereby the data transfer operation is initiated by one machine but involves data moving directly between two other machines. The latter capability is fundamental to Globus, because it allows for transfers to be initiated, monitored, and managed by the cloud-hosted Globus service.

Globus GridFTP incorporates fault recovery mechanisms that enable the resumption of data transfers upon failure, rather than a total restart---a crucial capability when transferring large datasets, where a failure can occur due to various network or system issues.
For security, Globus GridFTP supports strong authentication and data encryption to ensure that data transfers are secure and that only authorized users can access the service.
It also implements integrity checking to detect data corruption at any point along the path between source storage and destination storage.
(Data corruption while data at rest, while a concern \citep{bairavasundaram2008analysis}, is viewed as out of scope.)
In addition, its modular architecture allows for integration with a wide variety of storage systems, from conventional POSIX to high-performance parallel file systems and a range of object stores \citep{liu2021storage}.


The GridFTP protocol supports what we refer to here as \textit{chunked data transfer} by defining new commands such as SPAS (Striped Passive) and SPOR (Striped Port), and by adding stripe layout and block size options to the FTP RETR command \citep{allcock2003gridftp}.
The original Globus implementation of GridFTP \citep{allcock2005globus} implemented these commands, which enabled it to achieve $\sim$17~Gb/s of throughput between parallel file systems at NCSA and SDSC almost two decades ago.
In this \textit{server-side chunking} mechanism, the server determined how many GridFTP server processes to use for a chunked data transfer request. Though a powerful capability, a significant limitation is that the GridFTP control server must decide the number of nodes and processes to use for a chunked transfer, and in practice the GridFTP control server at one end had no knowledge about the configuration of the GridFTP server (e.g., number of DTNs) at the other end. This difficulty, plus some stability issues in the server-side chunking implementation, meant this mechanism was not widely used in practice.

From 2010 forward, Globus was re-architected as a hybrid architecture in which a cloud-hosted Globus service manages the activities of Globus Connect agents, for example by requesting pairs of such agents to perform file transfers in response to user requests \citep{foster2011globus}.
Today, tens of thousands of such agents are deployed on storage systems at thousands of institutions worldwide.
Globus service state is maintained in geographically replicated and thus highly reliable cloud storage, while file transfers proceed directly from one agent to another under the direction of the Globus service.
As we describe in the following, the Globus service's knowledge of Globus Connect agent configurations allows for the implementation of \textit{client-side chunking},
in which the Globus service, acting as a client to those agents, leverages partial file transfer mechanisms in the GridFTP protocol, in the form of ERET/ESTO (Extended Retrieve / Extended Store) commands \citep{allcock2003gridftp}, to drive chunked transfers.


\subsection{Previous Performance Evaluations}

Many studies of data transfer performance have been performed over the years, involving a wide variety of network environments, protocols, and workloads.
Here we comment just on some recent studies of Globus performance, as they provide context for subsequent discussion.

\citet{allcock2005globus} and \citet{ito2005parameter} studied the impact of concurrency and parallelism on achieved transfer performance. 
\citet{yildirim2012gridftp} conduct a more detailed investigation of the impact of GridFTP pipelining, parallelism, and concurrency on performance, and provide guidelines for setting these parameters. 
\citet{kettimuthu2015elegant} showed that by ensuring a sufficient, but not excessive, allocation of concurrency to the right transfers, overall performance of the resources can be improved significantly.
\citet{arslan2018big} present algorithms for dynamic adaptation of these parameters to improve performance.

\citet{liu2017explaining} applied machine learning methods to a large collection of Globus log data to estimate parameters for predictive models that yielded insights into factors determining end-to-end transfer performance.
One observation was that ``contention at endpoints can significantly reduce aggregate performance of even overprovisioned networks.''

\citet{liu2018cross} analyzed 40 billion GridFTP command logs totaling 3.3 exabytes and 4.8 million transfer logs collected by the Globus transfer service from 2014/01/01 to 2018/01/01.
Among many interesting observations, we note two:
First, they saw one integrity check failure per 1.26~TB, although admitting that the integrity checking protocol could not distinguish between true data corruption and a file changed deliberately during a transfer.
Second, they observed that most datasets transferred by Globus had only one file, and that 17.6\% of those datasets (or 11\% of the total) had a file size of $\ge$100 MB---motivating the need for distributing single-file transfers over multiple servers.

\citet{kettimuthu2018transferring} undertook a study in which they sought to move 1~PiB 
(2\textsuperscript{50}~B = 1.125~PB) in 24 hours over a 100~Gb/s network connecting Argonne National Laboratory and the National Center for Supercomputing Applications.
They succeeded ultimately in moving 1~PiB in 24~h 3~min without integrity checking (an average rate of 92.4 Gb/s) and in  30~h 52~min
(72 Gb/s) with integrity checking.
They achieved this rate via careful optimization of transfer parameters, including organizing the data to be transferred into 4~GB files and setting concurrency to 128 and parallelism to 1 (i.e., transferring 128 files concurrently, with a total of 128  TCP streams).

\citet{liu2019data} conducted a detailed study of per-file overheads in wide area Globus transfers.
Their results are applicable mainly to smaller files.

The Petascale DTN project \citep{dart2021petascale} was motivated, in the first instance, by the observation that despite HPC facilities being connected by 100~Gb/s networks and equipped with dedicated DTNs, achieved end-to-end data transfer performance among HPC facilities was often disappointing, rarely exceeding 10~Gb/s.
Thus in 2020 researchers at four such facilities (ALCF, NERSC, OLCF, and NCSA) undertook a systematic investigation of data transfer performance, with the goal of achieving routine transfer rates of 1~PB/week (a sustained 15~Gb/s) among the participating sites.
They used for these studies a $\sim$4.4~TB output dataset from a HACC simulation run \citep{HACCdata}, comprising a total of 211 directories and \num{19260} files, ranging in size from zero bytes to 11.3~GB.
As shown in \autoref{tab:petadtn}, performance among the four sites improved substantially over the course of the studies. The authors acknowledge multiple reasons for these improvements, from network upgrades to improved DTN hardware and changes to both the Globus implementation (e.g., see Section~IV.C in \citet{liu2018comprehensive}) and policies, and emphasize the importance of sustained monitoring to guide optimizations.


\begin{table}[h!]
\centering
\caption{Pairwise average data transfer rates, in Gb/s, reported by the Petascale DTN project \citep{dart2021petascale} between pairs of sites before (upper sub-table) and after (lower sub-table) optimization of DTN configurations.} 
\label{tab:petadtn}

\begin{tabular}{|l|c|c|c|c|}
\hline
& \multicolumn{4}{c|}{\textbf{Destination}}\\
\textbf{Source} & \textbf{ALCF} & \textbf{NCSA} & \textbf{NERSC} & \textbf{OLCF} \\ \hline
\hline
\multicolumn{5}{|c|}{\textbf{Performance at start of project}}\\
\hline
ALCF                 & -             & 13.4          & 10.0           & 10.5          \\ \hline
NCSA                 & 8.2           & -             & 6.8            & 6.9           \\ \hline
NERSC                & 7.3           & 7.6           & -              & 6.0           \\ \hline
OLCF                 & 11.1          & 13.3          & 6.7            & -   \\
\hline\hline
\multicolumn{5}{|c|}{\textbf{Performance at end of project}}\\
\hline
ALCF                & -             & 50.0          & 35.0           & 46.8          \\ \hline
NCSA                & 56.7          & -             & 22.6           & 34.7          \\ \hline
NERSC               & 42.2          & 33.7          & -              & 39.0          \\ \hline
OLCF                & 47.5          & 43.4          & 33.1           & -             \\ \hline
\end{tabular}
\end{table}

\section{Methodology}

Our major focus in this work is to enable rapid transfer of large files. To this end, we focus our attention on two goals.

Our first goal is to extend the Globus transfer service to orchestrate the actions of multiple data movers when moving large files, with the goal of achieving improved performance for large file transfer on POSIX file systems.  
With this mechanism, large files are chunked, and transferred, in parallel across multiple DTNs, from their source to their destination. We investigate the optimal chunk size capable of keeping all concurrent transfer sessions used, and incorporate changes to the transfer service to implement client-side chunking.
(We use the term chunking for this Globus capability here, rather than striping, so as to avoid confusion with striping as implemented in the Lustre file system, which we also discuss in the following.)

The second related goal is to optimize the process used by Globus to verify the integrity of transferred data, as the associated checksum computations and additional read operation have been shown to introduce significant costs in high-bandwidth transfers \citep{kettimuthu2018transferring}.

\begin{figure*}
    \centering
    \includegraphics[width=\textwidth]{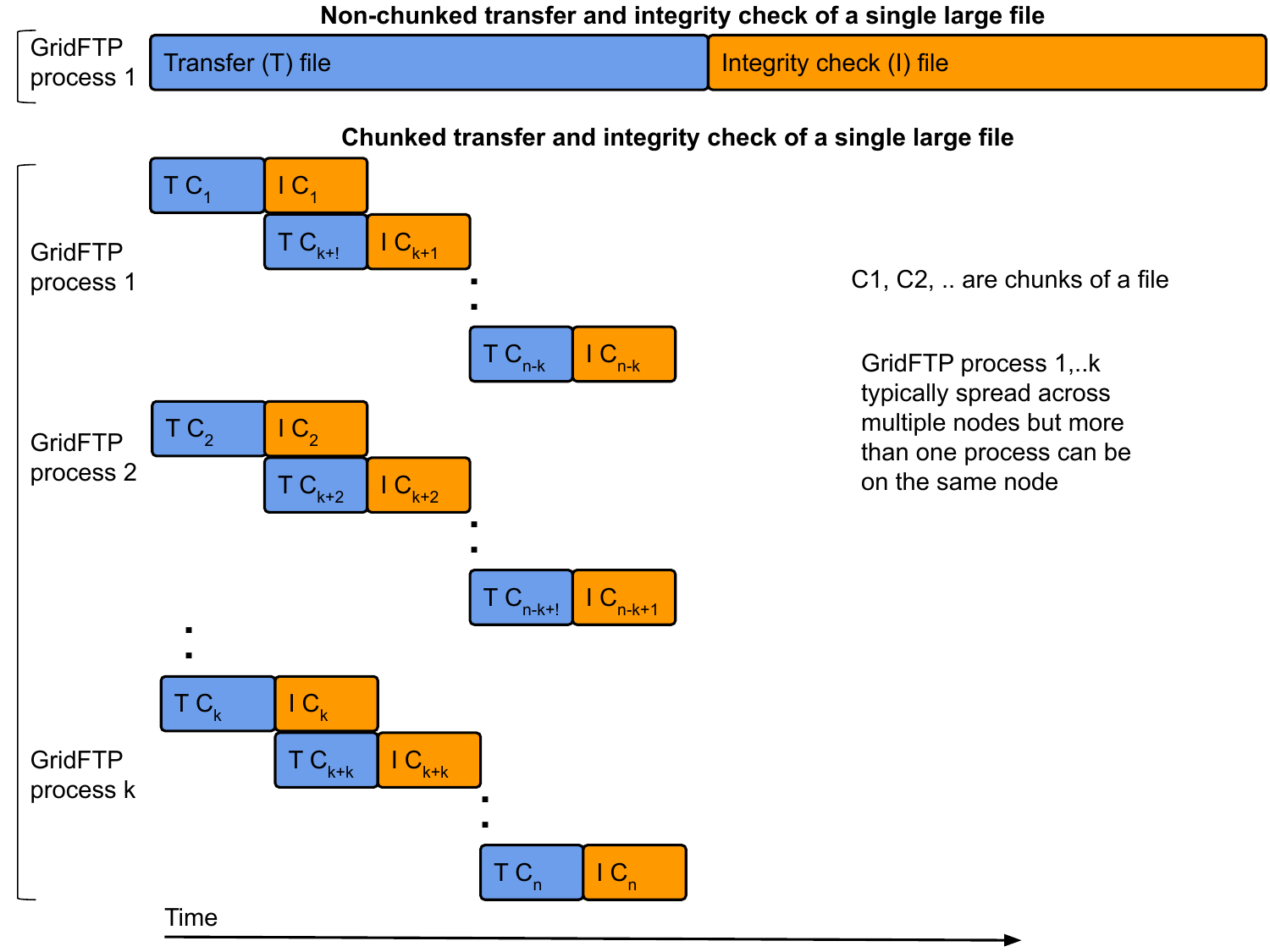}
    \caption{A sketch of activity over time for a non-chunked (above) and chunked (below) transfers, both with integrity checking. 
    With `time' on the horizontal axis,  the non-chunked transfer must wait until the entire file is transferred (blue) before performing its integrity check (orange), leading to longer end-to-end times. In the chunked file case, not only do multiple GridFTP processes transfer different portions of a file in parallel, but transfer and integrity checks execute concurrently. 
    (For simplicity, we show the integrity check cost being incurred only after the transfer; in practice, some modest cost also is incurred when first reading the file.)
    }
    \label{fig:rajcheck}
\end{figure*}

\subsection{Distribution of Transfers Over Multiple DTNs}

As described above, Globus leverages concurrency to transmit multiple files at one time, with the degree of concurrency supported by a particular Globus endpoint determined by configuration parameters that may be adjusted by the endpoint administrator.

Concurrency is a powerful accelerator of data transfers when many files are to be transmitted at the same time.
By allowing different data movers to proceed independently with reading and transmitting (or  receiving and writing), aggregate achieved I/O rates to the storage system increase, as do aggregate achieved network bandwidth in many cases---for example, when individual DTN connections to a border router have lower capacity than that router's connection to the wide area network.
Furthermore, as each data mover can operate on a separate file concurrently, no coordination costs are incurred. 
However, concurrency provides no benefits at all when transferring a single large file, as in that case just a single data mover will engage in data movement, leaving other data movers idle.
Whether transfer speed is limited by data mover file read/write performance or network send/receive performance \citep{liu2018cross}, the single data mover imposes a bottleneck.  
Similar concerns arise, albeit at a reduced level, when the number of large files is less than the optimal concurrency level for an endpoint.

This analysis suggests that a superior approach to transferring a single large file (or a small number of large files) could instead be to partition the task of transferring the large file(s) among multiple data movers.
To investigate the feasibility of this approach, we leveraged the partial file transfer capability of the Globus GridFTP server implementation to engage multiple data mover pairs at the source and destination to process disjoint file \textit{chunks} independently, reading and transmitting them at the source, and receiving and writing them at the destination.

In more detail, the transfer of a file with chunking proceeds as follows.
First, during the set up phase, we:
determine chunk size, $S$, and concurrency, $N$,  either via some heuristic or by reference to a configuration parameter;
create $N$ source-destination data mover pairs, and establish connections between the data movers in each pair; and
allocate chunks among data movers.
Then, each data mover pair proceeds to transmit chunks, using the extended retrieve (ERET) and extended store (ESTO) rather than the regular retrieve and store (RETR and STOR) commands, so as to allow for partial transfers. 
Pipelining is also used to ensure that individual data movers are not kept waiting for acknowledgements; as a consequence, chunks should not be too large. 
The implementation keeps track of which chunks have been transmitted successfully so as to enable efficient partial restarts upon failures.

\subsection{Optimization of Integrity Checking Calculations}

The Globus transfer service is configured by default to perform integrity checking on all files that it transfers, in order to detect data corruption due to such factors as faulty file system I/O or data transmission.  
Specifically, a Globus source node computes a 32-bit MD5 checksum for a file when reading it to transmit; the corresponding Globus destination node, upon receiving the file, first writes it to storage and then re-reads the file and computes a second checksum.
If the two checksums differ, an error is recorded and the file transfer is repeated.
Integrity checking errors are rare but do occur, often but not always in bursts due to faulty equipment, and thus this feature is an important element of the Globus service that, as far as we know, is rarely disabled. 
We note that Globus checksumming is in addition to and independent of checksumming performed by TCP, which only concerns data transmission, not file I/O, and furthermore uses an inadequate 16-bit checksum value \citep{stone2000crc}.

While essential for most if not all science applications, the costs of both computing the checksums and re-reading the file at the destination can be considerable.
Thus, we explored the feasibility of leveraging file chunking to enable concurrency and pipelining of checksum computations. 
Specifically, we extended the algorithm described above to compute and transmit a partial checksum with each ERET / ESTO pair, thus  distributing the costs of checksumming and performing the additional file read operation over multiple data movers, as illustrated in
\autoref{fig:rajcheck}.

\section{Experiments and Results}
\label{sec:results_weijian}


We conducted experiments between three HPC systems: the Argonne Leadership Computing Facility (ALCF) at Argonne National Laboratory;
the National Energy Research Scientific Computing Center (NERSC) at Lawrence Berkeley National Laboratory; and the Oak Ridge Leadership Computing Facility (OLCF) at Oak Ridge National Laboratory. 
Our experiments engaged high-performance Lustre file systems \citep{lustre} at each facility:
at ALCF, the Eagle file system; at NERSC, the Perlmutter scratch file system; and at OLCF, the Orion file system. 
Each facility operates multiple DTNs that are connected at high speeds to their associated file system, and to the ESnet wide area network at 100~Gb/s.

We performed five main sets of experiments, to measure:
\begin{itemize}
\item
The impact of Lustre striping on transfer performance.
\item 
The impact of Globus chunk size on transfer performance.
\item 
The impact of integrity checking on transfer performance.
\item 
The impact of varying the number of files being transferred on transfer performance.
\item 
The impact of chunking on transfers involving varying numbers of files.
\end{itemize}

\textit{All experiments were run with concurrency = 64 and parallelism = 4, and
all experiments that involved checksumming employed the default MD5SUM algorithm}.


\begin{figure}
    \centering
    \includegraphics[width=0.7\columnwidth]{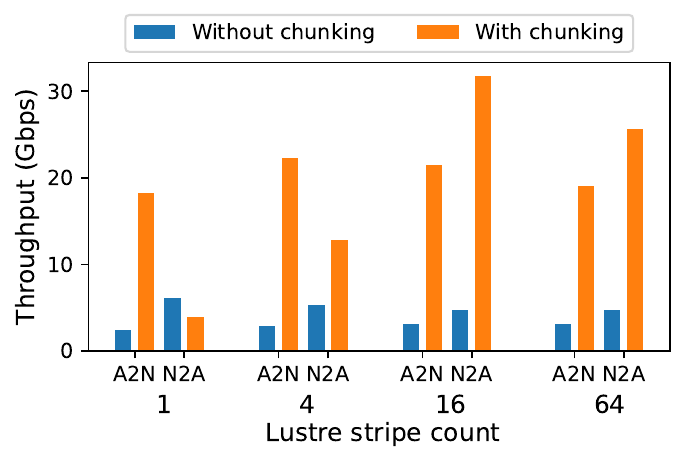}
    \caption{Impact of Lustre stripe count on Globus transfer performance for a 1$\times$2.5~TB file transfer between ALCF (A) and NERSC (N), both with and without chunking. All transfers were conducted
    without integrity checking.}
    \label{fig:lustre_test}
\end{figure}

\subsection{Striping in Lustre}

The Lustre file system is employed widely in HPC systems, including at the three facilities considered in this work.
We determined during our early investigations that a Lustre configuration parameter, specifically the number of Object Storage Targets (OSTs) over which a file was distributed, has a significant impact on the performance achieved when transferring one or a small number of large files. Here we report on experiments that quantify this impact and guide choices made in subsequent experiments.

Lustre allows administrators and users to specify striping of data across multiple OSTs, at the file system, directory, or individual file level.
Striped files can then be accessed concurrently
by multiple processes, boosting aggregate data throughput. 
Striping also enables the storage of files whose size exceeds the capacity of a single OST.

\def\MyHeightD{3.85cm}

\begin{figure*}
   \centering
\includegraphics[height=\MyHeightD]{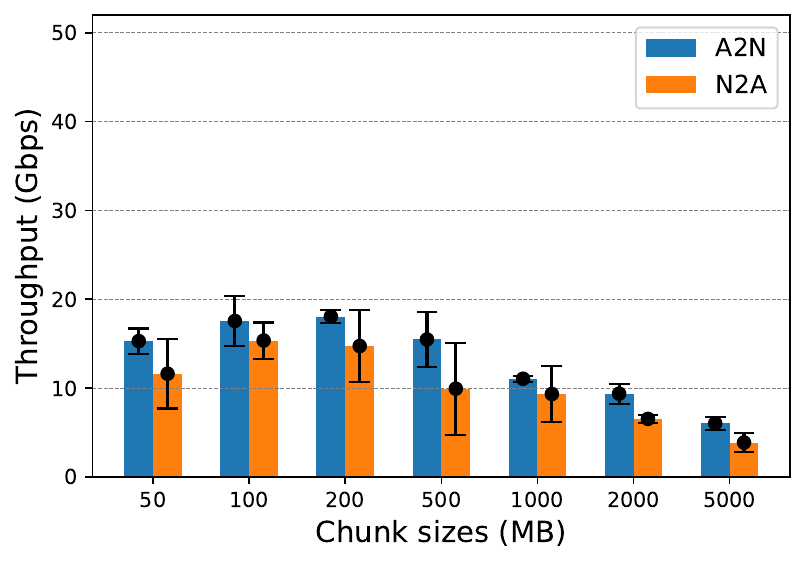}
\includegraphics[height=\MyHeightD]{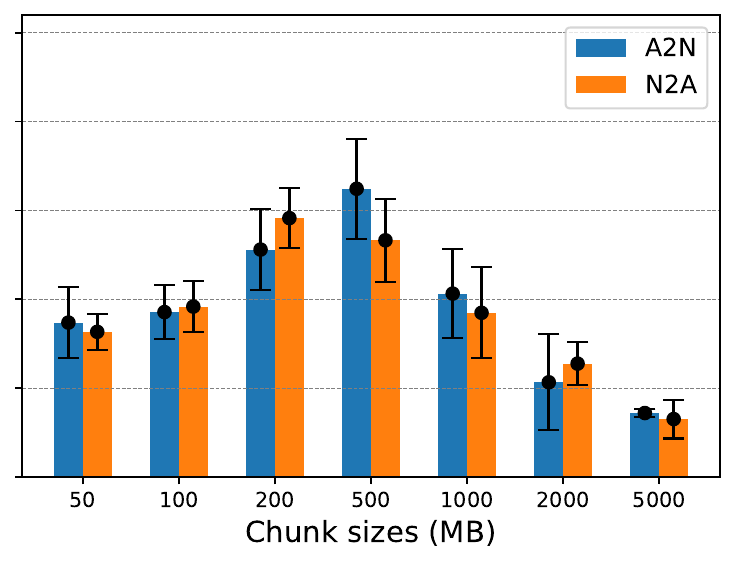}
\includegraphics[height=\MyHeightD]{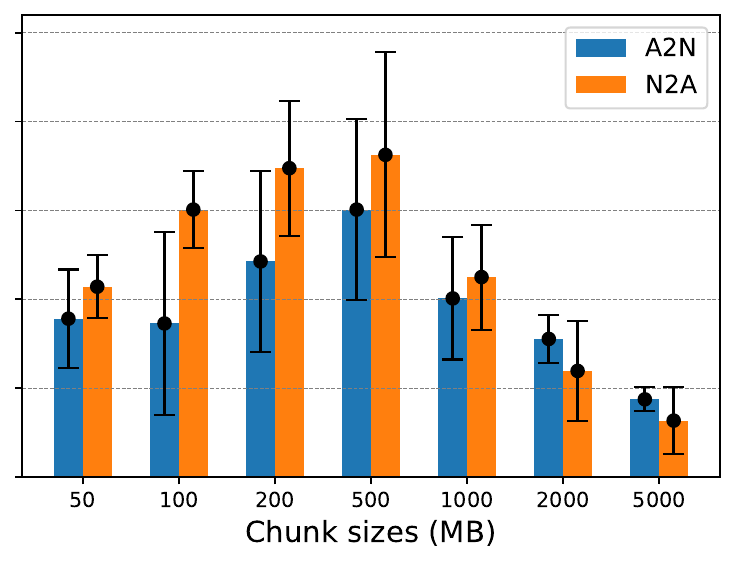}

{\small {\sffamily \hspace{0.4cm} 1$\times$500 GB \hspace{3.5cm} 5$\times$100 GB} \hspace{3.5cm} 20$\times$25~GB}

   \caption{\centering Impact of Globus chunk sizes on performance achieved for 500~GB transfers between ALCF and NERSC, for different numbers of files. All transfers were conducted with integrity checking.}
   \label{fig:chunk_sizes_A2N_N2A}
\end{figure*}

Striping also has drawbacks such as increased system overhead from heightened network activities and server competition, and can
increase the numbers of files corrupted by a single hardware failure. 
Lustre allows users to manage tradeoffs among these different factors by fine-tuning striping parameters, including stripe size and count, to optimize performance and reliability for specific needs---subject however to limits imposed by system administrators.

Unlike Globus chunking, which impacts both file I/O and network transfers, Lustre striping impacts file I/O alone.
Thus, for example, if transferring a chunked file with concurrency level of 2, then at the source endpoint two data movers work concurrently to read independent chunks of the file; if Lustre striping is engaged with striping 3, each data mover read operation retrieves data from three OSTs, and thus overall we would see read operations from the two data movers engaging up to six OSTs.
Thus, depending on configuration details, Globus chunking and Lustre striping can either complement each other or work at cross purposes.
Care must be taken to configure these different parameters in order to optimize performance and storage scalability.

Here we report on experiments in which we study the impact on data transfer performance of varying the Lustre stripe count, with the goal of determining the Lustre configurations that yield the highest transfer speeds.
We measure speeds achieved for a 1$\times$2.5~TB transfer with Lustre stripe counts from 1 to 16, both with and without chunking, and without integrity checking.
We perform transfer tests between ALCF and NERSC, in both directions---from ALCF to NERSC (A2N) and from NERSC to ALCF (N2A).
The Lustre stripe size is set to 1~MB.

Our results, in \autoref{fig:lustre_test}, show that in most test scenarios transfer speeds vary little (by less than 20\%) with Lustre stripe count.
However, for NERSC to ALCF transfers with chunking, increasing stripe count from 1 (the default on both systems) to 16 increased throughput by 8.1$\times$, from 3.92~Gb/s to 31.76~Gb/s. 
(It then declines for a stripe count of 64.)
We did not explore this phenomenon further, but observe that it highlights the importance of Lustre configuration for Globus transfers, which in many cases are rate-limited as much by file system I/O performance as by network bandwidth.
\textit{In all subsequent experiments, we set the Lustre stripe count to 16}.

\subsection{Impact of Globus Chunk Size} 

Chunk size, the size of each data segment transferred when using chunking, is a critical parameter for Globus transfers. 
To determine the impact of chunk size on Globus transfer throughput, we conducted tests between ALCF (A) and NERSC (N) for a variety of chunk sizes. We performed experiments for three 500~GB transfer scenarios, involving 1$\times$500 GB, 5$\times$100~GB, and 20$\times$25~GB files, respectively. 
We tested each configuration both from ALCF to NERSC (A2N) and NERSC to ALCF (N2A). 

We present our results in \autoref{fig:chunk_sizes_A2N_N2A}.
(In these results and those that follow, we ran each experiment 3--4 times and show in the figure both the average and one standard deviation either side of that average.)
In all three cases (1$\times$500~GB, 5$\times$100~GB, 20$\times$25~GB), performance tends first to \textit{increase} with chunk size over the range 50~MB to 500~MB and then to \textit{decrease} as chunk size increases further to 5000~MB.
We attribute the initial increasing trends to the larger chunk sizes allowing for more effective utilization of network bandwidth, and the subsequent decreases to decreased opportunities for parallelism for larger chunk sizes (e.g., with a chunk size of 5000~MB for a single 500~GB file, only 100 chunks are available, which is less than the product of the number of GridFTP control channel sessions and the number of separate TCP connections, which is  64$\times$4 = 256.)

We note that the performance gains from increasing chunk size are significantly less (at most 15\%) in the 1$\times$500~GB case. We attribute this result to greater bottlenecks and increased competition for resources when handling a single file. This observation is underscored by the fact that the throughput improvement for the 20$\times$25~GB case is significantly greater than that with 5$\times$100~GB files.



\def\MyHeightC{3.6cm}

\begin{figure*}
   \centering
        \includegraphics[height=\MyHeightC]{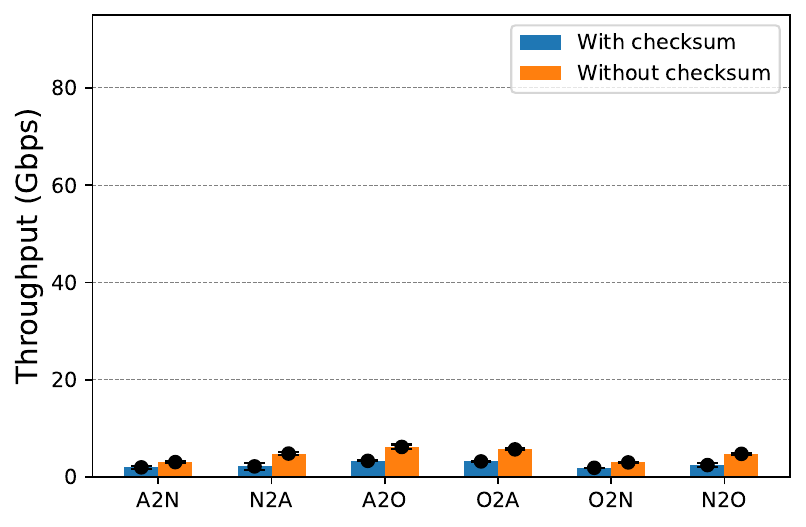}
        \includegraphics[height=\MyHeightC]{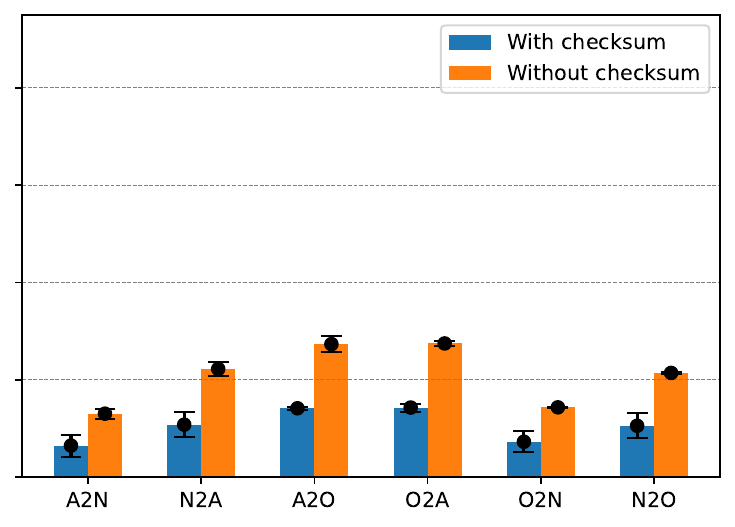}
        \includegraphics[height=\MyHeightC]{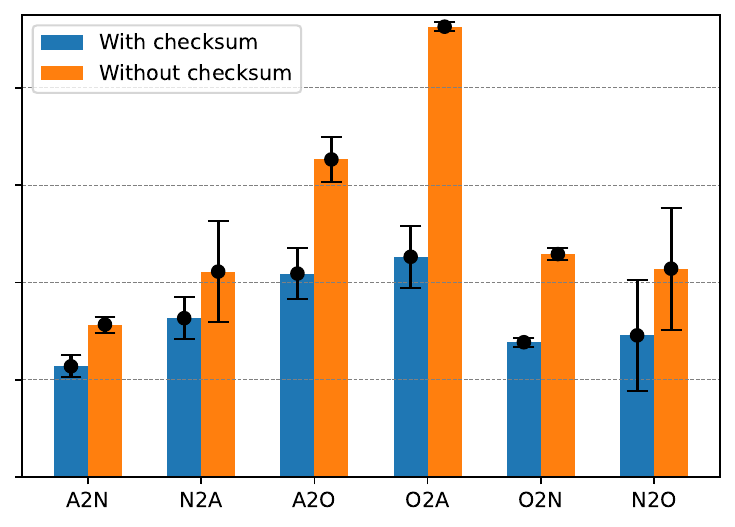}
       \includegraphics[height=\MyHeightC]{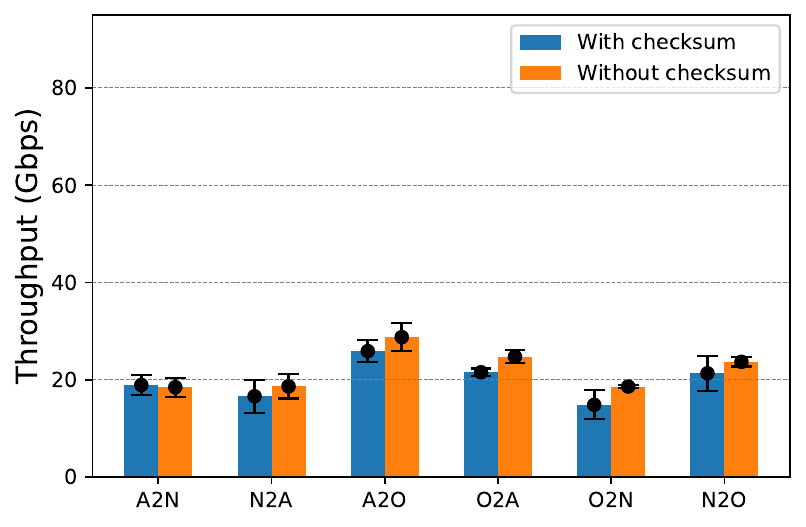}
       \includegraphics[height=\MyHeightC]{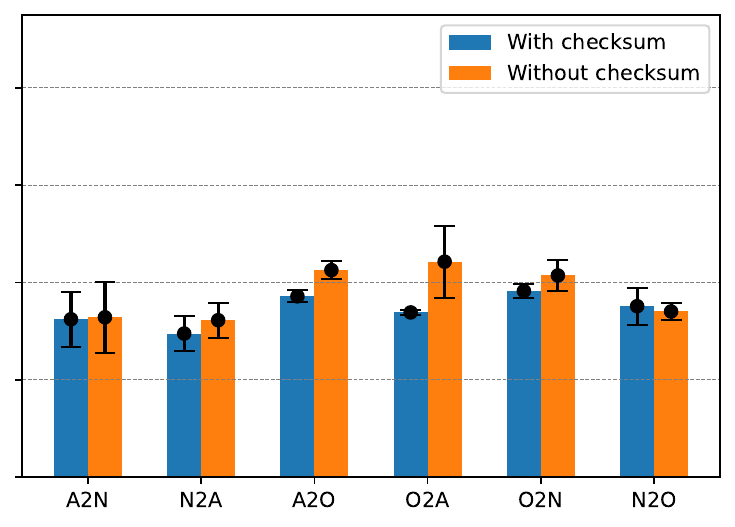}
       \includegraphics[height=\MyHeightC]{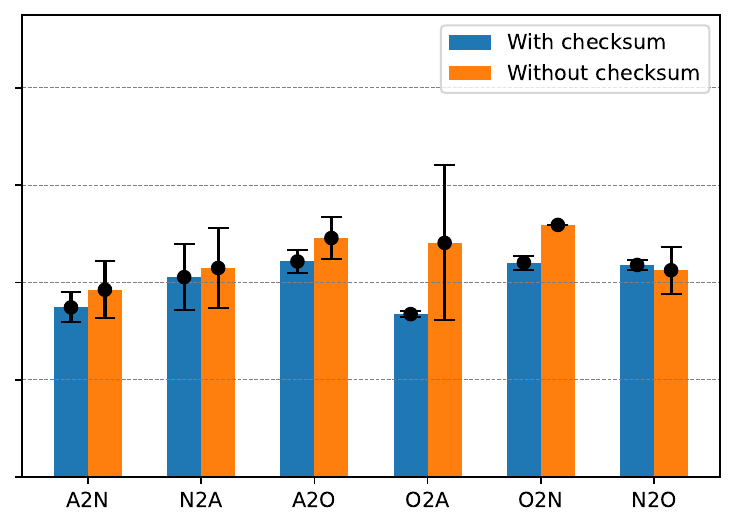}

        {\small {\sffamily \hspace{0.5cm} 1$\times$500 GB \hspace{3.5cm} 5$\times$100 GB} \hspace{3.5cm} 20$\times$25~GB}

       \caption{\centering 500~GB transfers, in from 1 to 20 files, among three facilities, with and without integrity checking. 
       \textbf{Above}: Without chunking.
       \textbf{Below}: With chunking.}
   \label{fig:checksum_striping}
\end{figure*}

\begin{figure}
    \centering
    \includegraphics[width=0.7\columnwidth]{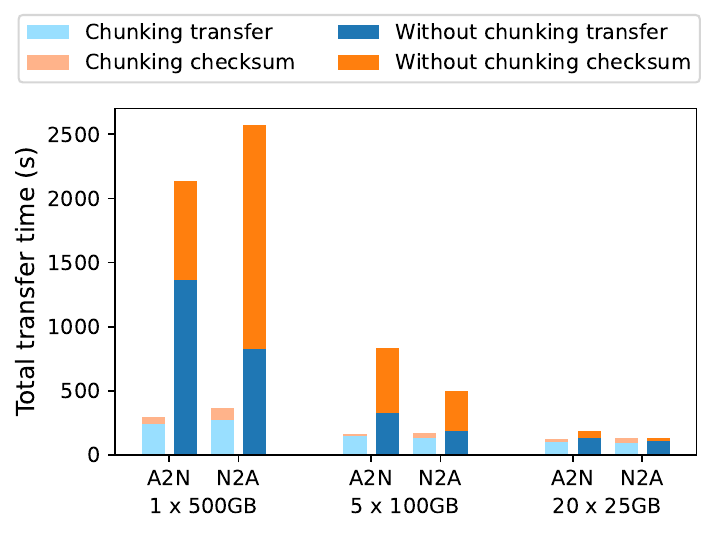}
    \includegraphics[width=0.7\columnwidth]{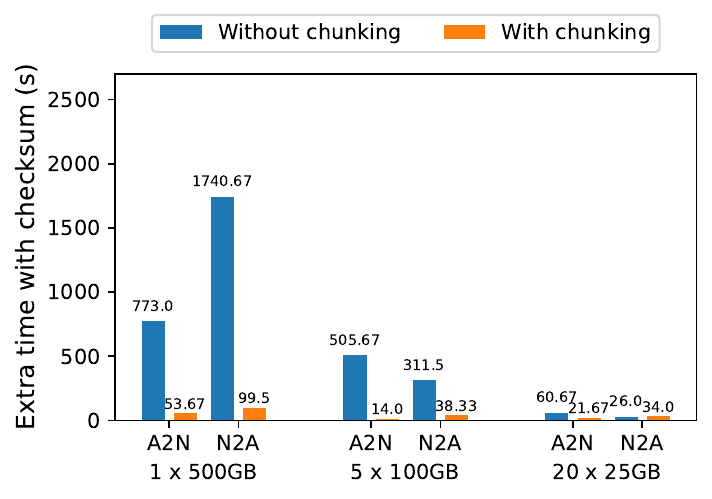}
    \caption{Average transfer and integrity check times for three different 500~GB transfer tasks between ALCF and NERSC, with and without chunking.  
    \textbf{Above}: Stacked transfer and integrity check times.
    \textbf{Below}: Integrity check times only.
    }
    \label{fig:checksum_time}
\end{figure}

\subsection{Integrity checking Costs}

We emphasize that in most cases, Globus integrity checking is crucial 
for ensuring correct data tranmission.
Nevertheless, we want to understand the performance impact of integrity checking so as to guide optimizations.
To that end, we conducted tests across ALCF, NERSC, and OLCF in which we measured data transfer performance for the same three tasks considered in the chunk size study (i.e., 1$\times$500~GB, 5$\times$100~GB, and 20$\times$25~GB files), with and without integrity checking. For the chunking experiments, we experimented with different chunk sizes and selected the configuration that delivered the fastest throughput, which in all subsequent experiments was either 200~MB or 500~MB.

We show in \autoref{fig:checksum_striping} throughput with and without integrity checking, both without (above) and with (below) chunking.
We observe first (upper subfigures) that integrity checking impacts throughput significantly in the non-chunking cases.
For example, transfer speed is roughly halved for both A2N and N2A, an effect that declines somewhat for some source-destination pairs with more files, but remains pronounced.
With chunking (lower subfigures), the performance degradation persists but is much less pronounced, particularly when more files are involved.

To focus in on these differences, we show in \autoref{fig:checksum_time} a different view of the A2N and N2A data.
Transfer and integrity checking (`checksum') times, averaged across the sets of 4 experiments for which results are provided in \autoref{fig:checksum_striping}, are presented in stacked bar chart form, both with and without chunking. 
We observe, looking left to right, that as the number of files increases from 1 to 20, transfer and integrity checking times both decrease substantially.
For example, for transfer tasks \textit{without} chunking from ALCF to NERSC, the average integrity checking time decreases significantly, from 773~s to 60.7~s, as the number of files increases from 1 to 20.
\textit{With} chunking, the decrease is from 53.7~s to 21.7~s.
For the 1$\times$500~GB task, integrity checking times without and with chunking are 773~s and 53.7~s, respectively, emphasizing the importance of chunking for single (or few) large file(s) transfers.
Overall, we see that by allowing integrity checking operations to be performed in parallel, chunking enhances throughput significantly. 
These results underscore the benefits of parallelizing integrity checking for data transfer performance. 


\def\MyHeightA{3.6cm}

\begin{figure*}
   \centering
       \includegraphics[height=\MyHeightA]{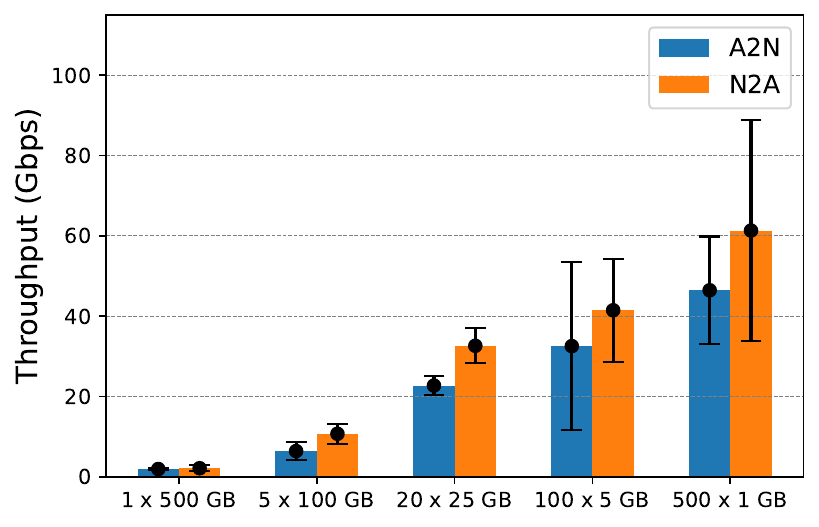}
       \includegraphics[height=\MyHeightA]
       {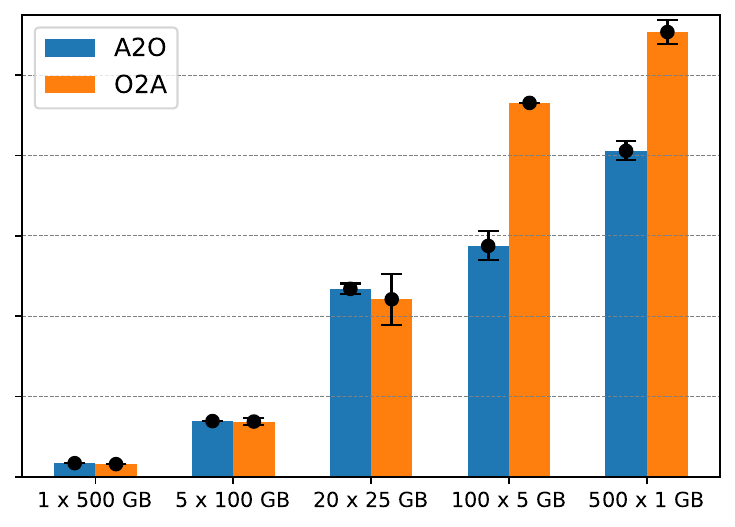}
       \includegraphics[height=\MyHeightA]
       {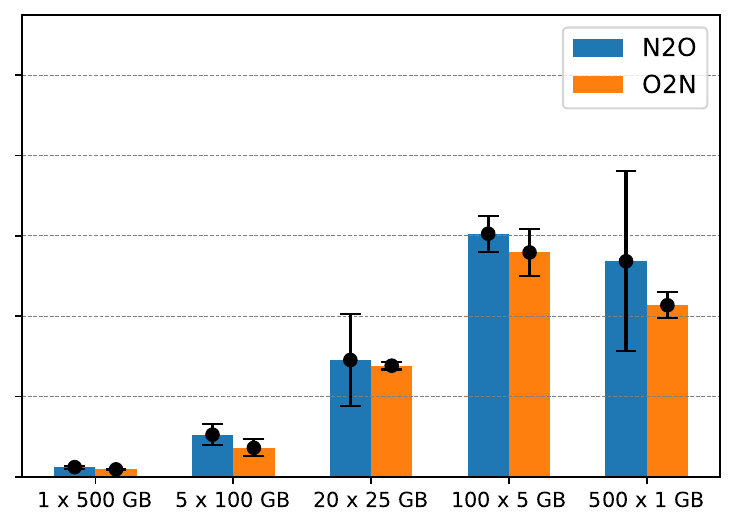}
       \includegraphics[height=\MyHeightA]{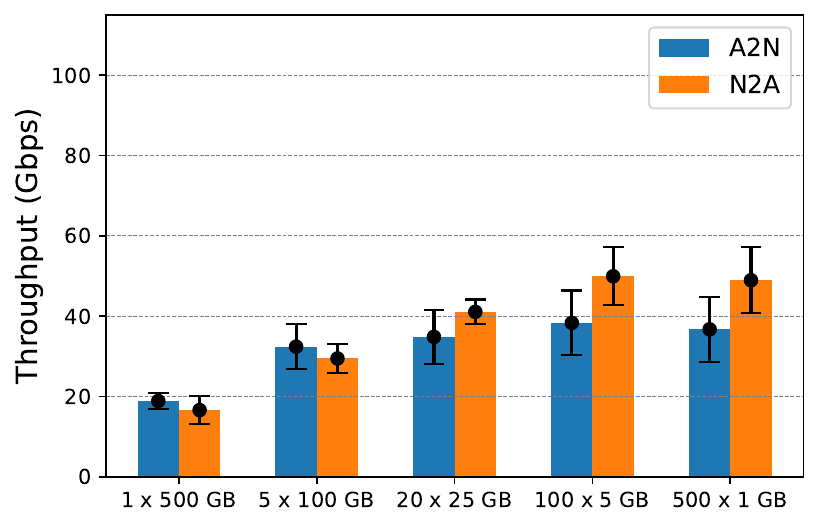}
       \includegraphics[height=\MyHeightA]
       {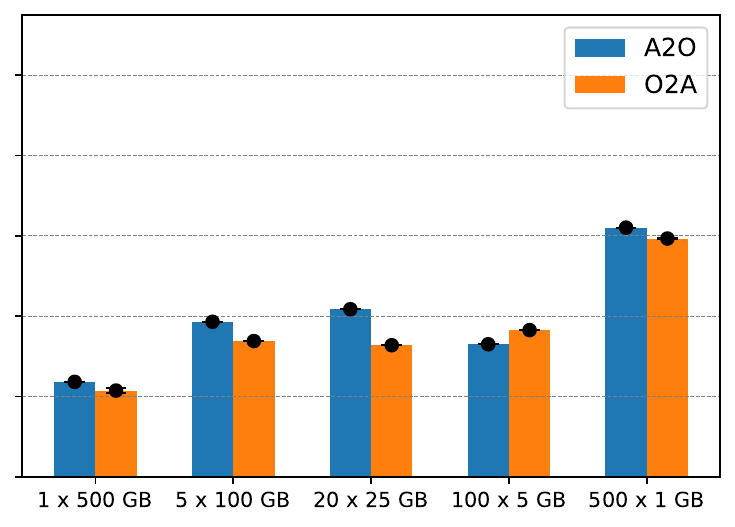}
       \includegraphics[height=\MyHeightA]
       {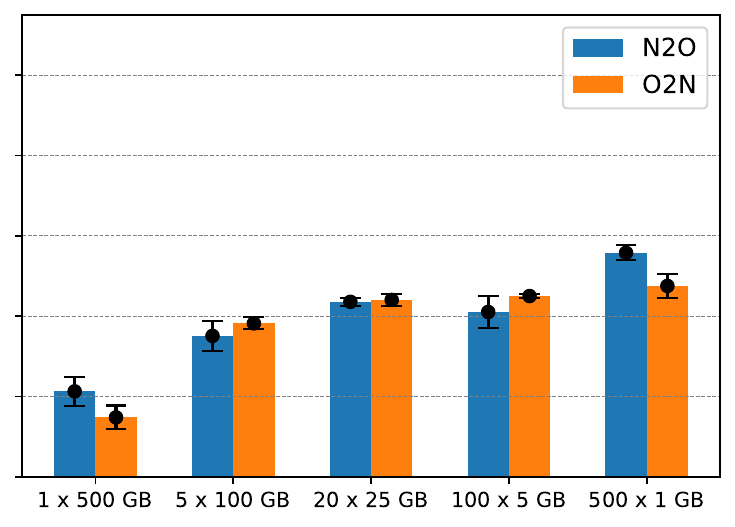}

       {\small {\sffamily \hspace{0.5cm} A2N and N2A \hspace{2.5cm} A2O and O2A} \hspace{2.5cm} N2O and O2N}
       
   \caption{\centering 500~GB transfers, of from 1 to 500 files, among different pairs of three facilities.
   \textbf{Above}: Without chunking. \textbf{Below}: With chunking. Integrity checking is enabled for all tasks}
   \label{fig:single_multiple}
\end{figure*}

\def\MyHeightB{3.6cm}

\begin{figure*}
   \centering
       \includegraphics[height=\MyHeightB]{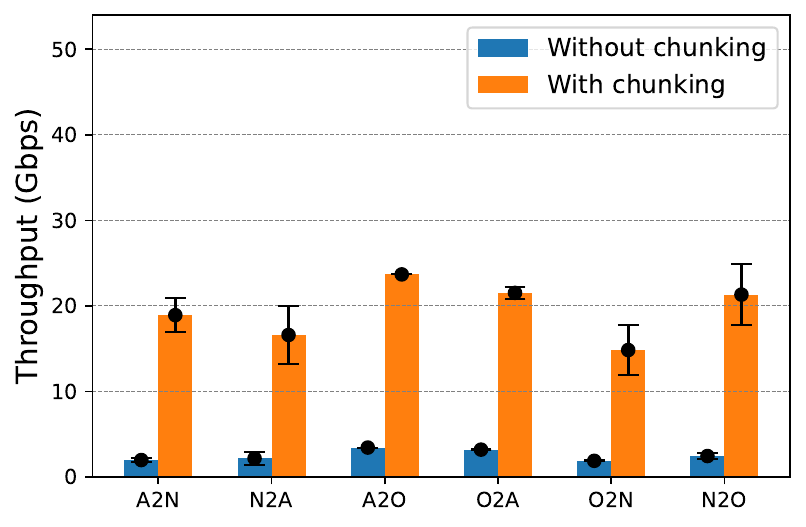}
       \includegraphics[height=\MyHeightB]{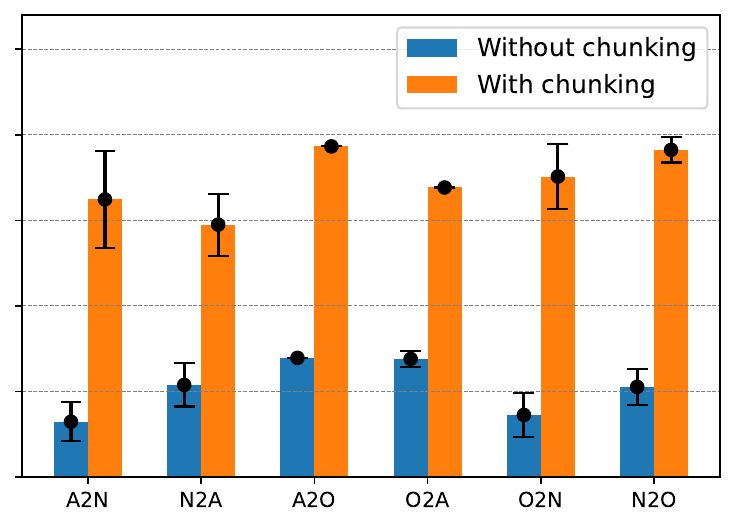}
       \includegraphics[height=\MyHeightB]{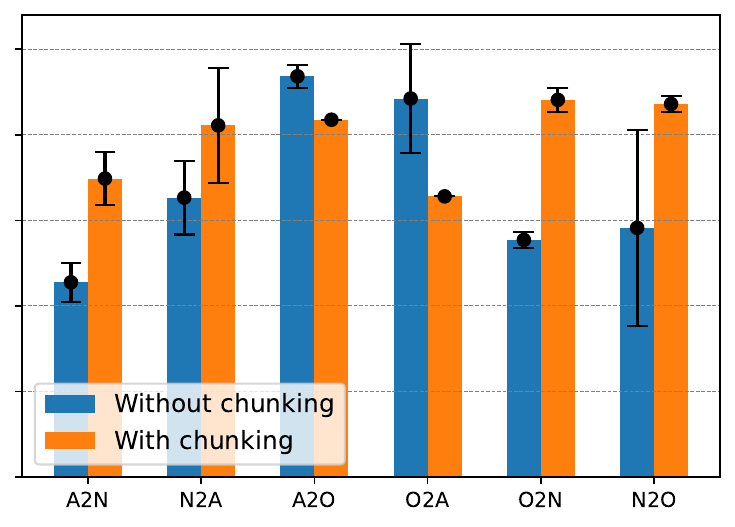}

    {\small {\sffamily \hspace{0.5cm} 1$\times$500 GB \hspace{3.5cm} 5$\times$100 GB} \hspace{3.5cm} 20$\times$25~GB}

   \caption{\centering Transfer speed for 500~GB data, in 1$\times$500~GB, 5$\times$100~GB, and 20$\times$25~GB, among the three facilities, with and without chunking.
   Integrity checking is enabled for all tasks.}
   \label{fig:striping_no_striping_wo_checksum}
\end{figure*}

\subsection{Single vs.\ Multiple File Transfers}

Here we investigate how performance varies with the number of files in a task. 
In addition to the 1$\times$500~GB, 
5$\times$100~GB, and
20$\times$25~GB tasks considered in previous experiments, we also consider
100$\times$5~GB
and 500$\times$1~GB.
We measured throughput both with and without chunking, and with integrity checking. 
For the tests that employ chunking, we also assessed multiple chunk sizes to determine the configuration that yields optimal performance.

Our results, in \autoref{fig:single_multiple}, demonstrate that for this fixed total transfer size of 500~GB, it is always faster to transfer multiple smaller files than a single large file, due to the considerable opportunities for parallelism in the former case, although the magnitude of this difference is reduced when chunking is employed.
For example, for ALCF to NERSC \textit{without} chunking (the upper row in the figure), increasing the number of files from 1 to 500 boosted transfer speed from 1.98 Gb/s to 46.48 Gb/s: a 23-fold increase. Similarly, for NERSC to ALCF, increasing the file count to 500 results in more than a 28-fold speedup. 
However, when Globus chunking is enabled (the bottom row), the performance differential between single-file and many-file transfers decreases significantly. For the ALCF to NERSC route, the speedup reduces from 23  to 1.9, and for NERSC to ALCF, from 28 to 3.

\subsection{Transfers With and Without Chunking}

Finally, we explore the impacts of chunking on transfer tasks involving 1$\times$500~GB, 5$\times$100~GB, and 20$\times$25~GB files.
Our results, in \autoref{fig:striping_no_striping_wo_checksum}, show that chunking has clear benefits when transferring a few files,
but that these benefits largely disappear for many files.
Specifically, we find that chunking yields peak speedups of up to 9.5$\times$ (for A2N) and 5$\times$ (for A2N) for the single-file and five-file transfer cases, respectively, 
but for the 20-file case, the maximum speedup diminishes to 1.6 (O2N), while there is even some slowdown in some cases (e.g., O2N).
It may be significant that chunking performs better in the 20-file case for the larger round-trip-time (RTT) ALCF-NERSC and NERSC-OLCF cases than for the lower RTT ALCF-OLCF case.

\section{Related Work}

We summarized above several studies that focused particularly on Globus GridFTP performance. Here we note other relevant work.

\citet{tierney1994distributed,tierney1999network} conducted pioneering work on striping as a means of accelerating remote data access applications, and \citet{kettimuthu2010lessons} reviewed factors that can influence the speed of large transfers.

The performance achieved for a particular data transfer can depend significantly on numerous configuration choices, including the network protocol used: e.g., TCP or UDP-based alternatives \citep{he2002reliable,gu2007udt}; number of data movers \citep{kettimuthu2014modeling}; number of TCP streams \citep{sivakumar2000psockets,hacker2002end,lu2005modeling}; TCP window size \citep{hstcp}; TCP variant \citep{bullot2003evaluation,leith2004h,wei2006fast}; degree of pipelining, file system striping; and use of redundant paths \citep{zhang2004transport}. Researchers have investigated the impact of such parameters on the performance achieved for different transfer tasks and in different environments \citep{ito2005parameter}, and proposed methods for selecting such parameters automatically \citep{prasad2003socket,yildirim2015application,arslan2018high}.
They have also investigated the impact of transfer parameters on different performance metrics (e.g., latency vs.\ bandwidth) and on properties other than performance, such as energy consumption \citep{alan2015energy} and impact on competing flows \citep{hacker2002end,lu2005modeling}. 
These are factors that could be considered in Globus, which currently focuses on bandwidth.

In other related work, \citet{liu2016towards} discuss block-level streaming computation of checksums to accelerate integrity checking, while
\citet{arslan2018low} discuss ways in which integrity checking costs can be reduced by careful organization of checksum computations and file I/O operations.
\citet{arifuzzaman2021online} use online optimization to select transfer parameters.
\citet{charyyev2020riva} examine how detection of file corruption errors can be enhanced by ensuring that checksum calculations are performed on disk-resident rather than cached data.
Various researchers have investigated compression of data to be communicated over networks \citep{cappello2019use,foster2017computing}.

\section{Conclusions}

We have reported on our development and evaluation of a new capability in the Globus transfer service designed to accelerate movement of individual large files.
The key development here is the addition of support for the (logical) chunking of large files into disjoint subsets that are then transmitted by distinct data movers, in ways that also allow for enhanced overlapping of checksum computations with data movement.
We demonstrate by careful experimentation that these developments can deliver significant performance benefits.
For example, we find that when transferring a single 500GB file from ALCF to NERSC, chunking increases performance by a factor of 9.5.

We can also point to other opportunities for further optimizations.
In the current implementation, chunking is enabled manually, either on a per-user basis or by a user labeling a transfer.
Large-scale deployment would likely require automation of decisions concerning which transfers to chunk and what chunk size to employ, with the latter potentially being set based on the file in hand.
Our results also suggest that significant opportunities remain for further optimization of integrity checking, perhaps by application of methods proposed by \citet{arslan2018big}.

This work also points to the importance of managing parallel file system striping parameters. 
We found that for transfers between Lustre file systems, tuning the Lustre stripe count can improve the transfer throughput by up to 8.1$\times$.
Thus we may want Globus to allow the administrator of a Globus collection that contains a few large files to enable chunking for that dataset. The specified stripe width would then need to be communicated from the Globus service to the local Globus Connect Server agent, which in turn would set the stripe width using the API or CLI provided by the file system.

Our experiments also provide insights into the growing costs of integrity checking as increased network speeds reduce the time taken for data transmission. Such trends may motivate the use of alternative checksumming algorithms.

\section*{Acknowledgements}
We acknowledge support from the Exascale Computing Project (17-SC-20-SC), a collaborative effort of the U.S.\ Department of Energy Office of Science and the National Nuclear Security Administration.
We are grateful to the ALCF, NERSC, and OLCF, DOE Office of Science User Facilities supported under Contracts DE-AC02-06CH11357, DE-AC02-05CH11231, and DE-AC05-00OR22725, respectively, for access to computing resources used in experiments.

\balance

\bibliographystyle{plainnat}
\bibliography{refs.bib}

\end{document}